\documentclass[a4paper,fleqn,usenatbib]{mnras}

\usepackage{newtxtext,newtxmath}
\usepackage[T1]{fontenc}
\usepackage{ae,aecompl}

\usepackage{float}

\usepackage{graphicx}	% Including figure files
\usepackage{amsmath}	% Advanced maths commands
\usepackage{amssymb}	% Extra maths symbols
\usepackage{lineno}
\usepackage{afterpage}
\title[Formation of Ellipticals in Groups]{Galaxy and Mass Assembly (GAMA): Formation and Growth of Elliptical Galaxies in the Group Environment}

\author[S. Deeley et al.]{
Simon Deeley,$^{1,2}$\thanks{E-mail: simon.deeley@uqconnect.edu.au}
Michael J. Drinkwater,$^{1,2}$
Daniel Cunnama,$^{3,4}$
\newauthor Joss Bland-Hawthorn,$^{5}$
Sarah Brough,$^{6}$
Michelle Cluver,$^{7}$
Matthew Colless,$^{8}$
\newauthor Luke J. M. Davies,$^{9}$
Simon P. Driver,$^{10,11}$
Caroline Foster,$^{6}$
Meiert W. Grootes,$^{12}$
\newauthor A. M. Hopkins,$^{6}$
Prajwal R. Kafle,$^{13}$
Maritza A. Lara-Lopez,$^{14}$
Jochen Liske,$^{15}$
\newauthor Smriti Mahajan,$^{16}$
Steven Phillipps,$^{17}$ 
Chris Power,$^{2,13}$
 Aaron Robotham.$^{13}$
\\
$^{1}$School of Mathematics and Physics, University of Queensland, Brisbane, Queensland 4072, Australia\\
$^{2}$ARC Centre of Excellence for All-Sky Astrophysics (CAASTRO)\\
$^{3}$South African Astronomical Observatory, PO Box 9, Observatory, Cape Town 7935, South Africa\\
$^{4}$Department of Physics and Astronomy, University of the Western Cape, Cape Town 7535, South Africa\\
$^{5}$Sydney Institute for Astronomy, School of Physics A28, University of Sydney, NSW 2006, Australia\\
$^{6}$Australian Astronomical Observatory, PO Box 915, North Ryde, NSW 1670, Australia\\
$^{7}$Department of Physics and Astronomy, University of the Western Cape, Robert Sobukwe Road, Bellville, 7535, South Africa\\
$^{8}$Research School of Astronomy and Astrophysics, Australian National University, Canberra, ACT 2611, Australia\\
$^{9}$International Centre for Radio Astronomy Research, University of Western Australia, Perth, Western Australia 6009, Australia\\
$^{10}$The University of Western Australia, M468, 35 Stirling Highway, Crawley, WA 6009, Australia\\
$^{11}$SUPA, School of Physics and Astronomy, University of St Andrews, North Haugh, St Andrews, UK, KY16 9SS\\
$^{12}$ESA/ESTEC SCI-S, Keplerlaan 1, 2200 AG Noordwijk, Netherlands\\
$^{13}$International Centre for Radio Astronomy Research, University of Western Australia, Perth, Western Australia 6009, Australia\\
$^{14}$Instituto de Astronomía, Universidad Nacional Autónoma de México, A.P. 70-264, 04510 México, D.F., México\\
$^{15}$Hamburger Sternwarte, Universit{\"a}t Hamburg, Gojenbergsweg 112, 21029 Hamburg, Germany\\
$^{16}$Indian Institute of Science Education and Research Mohali- IISERM, Knowledge City, Sector 81, SAS Nagar, Manauli, PO 140306, INDIA\\
$^{17}$Astrophysics Group, School of Physics, University of Bristol, Bristol BS8 1TL, UK\\
}

\date{Accepted XXX. Received YYY; in original form ZZZ}

\pubyear{2016}

\begin{document}
%\linenumbers
\label{firstpage}
\pagerange{\pageref{firstpage}--\pageref{lastpage}}
\maketitle

% Abstract of the paper  NOTE 250 word limit ****
\begin{abstract}
There are many proposed mechanisms driving the morphological transformation of disk galaxies to elliptical galaxies. In this paper, we determine if the observed transformation in low mass groups can be explained by the merger histories of galaxies. We measured the group mass-morphology relation for groups from the Galaxy and Mass Assembly group catalogue with masses from 10$^{11}$ - 10$^{15}$ M$_{\odot}$. Contrary to previous studies, the fraction of elliptical galaxies in our more complete group sample increases significantly with group mass across the full range of group mass. The elliptical fraction increases at a rate of 0.163$\pm$0.012 per dex of group mass for groups more massive than 10$^{12.5}$ M$_{\odot}$. If we allow for uncertainties in the observed group masses, our results are consistent with a continuous increase in elliptical fraction from group masses as low as 10$^{11}$M$_{\odot}$. We tested if this observed relation is consistent with merger activity using a GADGET-2 dark matter simulation of the galaxy groups. We specified that a simulated galaxy would be transformed to an elliptical morphology either if it  experienced a major merger or if its cumulative mass gained from minor mergers exceeded 30 per cent of its final mass. We then calculated a group mass-morphology relation for the simulations. The position and slope of the simulated relation were consistent with the observational relation, with a gradient of  0.184$\pm$0.010 per dex of group mass. These results demonstrate a strong correlation between the frequency of merger events and disk-to-elliptical galaxy transformation in galaxy group environments.
\end{abstract}


%%%%%%%%%%%%%%%%%%%%%%%%%%%%%%%%%%%%%%%%%%%%%%%%%%

%%%%%%%%%%%%%%%%% BODY OF PAPER %%%%%%%%%%%%%%%%%%

%1
\section{Introduction}

After making the first observations of the different galaxy morphologies \citep{1926ApJ....64..321H}, Hubble placed them into two main categories, ellipticals and spirals. His naming these classes as `early-type' and `late-type' respectively has been taken to imply that galaxies progressively evolved from ellipticals into spirals. Subsequent research \citep[e.g.][]{1984ApJ...285..426B,1997ApJ...490..577D} revealed that the relative fraction of elliptical to spiral galaxies decreases with increasing redshift, and it is now widely believed that elliptical galaxies have evolved from spirals. What physical processes are driving these transformations remains uncertain, and is a major area of active research. 

The prevalence of large elliptical galaxies within high density environments relative to the field is well known. \citet{1980ApJ...236..351D} investigated 55 high mass galaxy clusters and showed that the elliptical and S0 fractions increase with the increasing projected cluster density. Enhanced elliptical fractions relative to the field have also been seen in smaller galaxy groups \citep{2006MNRAS.370.1223B}. The elliptical fraction increases with both the group's X-ray luminosity and velocity dispersions \citep{2006MNRAS.370.1223B}, both of which are proxies of the group mass. What causes these enhanced elliptical fractions in high density environments, in particular the role played by merger activity, remains a subject of debate. 

Merger activity has long been considered as a possible mechanism for the formation of elliptical galaxies \citep[e.g.][]{1972ApJ...178..623T,1977egsp.conf..401T}. Observations have revealed that the majority of galaxies with masses $>10^{10} M_{\odot}$ have experienced 1-2 major merger events within $z<1.2$ \citep{2009MNRAS.394.1956C}.
After assuming a major merger between two gas-rich disk galaxies forms a quenched elliptical, \citet{2008ApJS..175..390H} showed that expected major merger rates account for the observed fraction of red ellipticals as a function of redshift. Alternatively, others \citep[e.g.][]{2010ApJ...721..193P} argue internal feedback mechanisms and non-dynamical features in higher density environments are responsible for the transformations.  

 Computer simulations have added support to the formation of ellipticals via mergers. \citet{2013ApJ...778...61T} simulated mergers within groups of halo mass $10^{11} - 10^{13}M_{\odot}$ containing 3-25 spiral galaxies, and found that the resulting central galaxies had S\'ersic profiles matching those of ellipticals.  \citet{2007A&A...476.1179B} determined that the critical factor in what impact a series of mergers has on galaxy morphology is the cumulative mass ratio. A series of minor mergers can have the same impact as a single major merger, and so they should be taken into account when considering the impact of mergers on galaxy transformation.

The environment within large galaxy clusters is very rich and complex, with varying dynamical interactions and an intracluster medium of hot, turbulent X-ray gas \citep{2014Natur.515...85Z}, leading to many complicating processes which may be contributing to galactic evolution. Low mass galaxy groups feature a far simpler environment, lessening or eliminating many theorised transformation processes. As such groups likely retain enhanced merger activity due to their above-field densities, merger activity remains as a possible dominant driving force behind galaxy evolution, making galaxy groups ideal environments to test how merger activity can influence galaxy evolution. 

\citet{2012MNRAS.423.3478H}  and \citet{2009MNRAS.393.1324B} took advantage of the Galaxy Zoo project and the Sloan Digital Sky Survey (SDSS) C4 group catalogue \citep{2005AJ....130..968M} to construct group mass-morphology relations down to a group halo mass of 10$^{13}M_{\odot}$. Both concluded from their analysis that there is little variation of the elliptical fraction with halo mass. However, the C4 catalogue is only complete for halo masses above $10^{14.7}M_{\odot}$ \citep{2012MNRAS.423.3478H}, and they also limited their study to galaxies with stellar masses > $10^{10} M_{\odot}$ (Hoyle et al.) and > $10^{9.8} M_{\odot}$ (Bamford et al.). Elliptical galaxies tend to have higher stellar masses than spirals, and hence smaller spiral galaxies in lower mass groups are less likely to be detected compared to their elliptical neighbours. A more complete determination of the elliptical fraction against group mass relationship, and subsequent comparison with merger activity derived from simulations, would lead to a clearer insight not only into galaxy evolution within these environments, but also help shed light on the extent to which merger activity is responsible for morphological transformations in general.
 
There is evidence for merger activity even in the limit of the smallest `groups', galaxy pairs. \citet{Scudder2012} measured star formation rates in galaxy pairs from the SDSS survey. Galaxies experiencing mergers had significantly enhanced star formation, although the strongest effect was seen in major mergers. This is associated with population changes as the red fraction of galaxies in SDSS pairs is higher than that of a control sample \citep{Patton2011}. There is also evidence from the GAMA survey that the effect of this environment on the pair galaxies depends on their mass. \citet{Davies2015} measured the effect of close interactions: star formation in the lower mass galaxy is suppressed, while it increases in the higher mass galaxy.  \citet{Robotham2014} examined the mass growth in GAMA pair galaxies. Star formation dominates mass growth in the smaller galaxies and merger events dominate mass growth in larger galaxies.
 
The aims of our study are first to determine the mass-morphology relation observed across small group masses down to 10$^{11}M_{\odot}$, and second, to determine to what extent this relationship can be explained by merger activity. The first aim is achieved using group catalogue datasets \citep{2011MNRAS.416.2640R} from the Galaxy and Mass Assembly (GAMA) survey \citep{2011MNRAS.413..971D}. To test the merger hypothesis, we compare these observational results to a merger-history derived mass-morphology relation created from the results of a GADGET-2 dark-matter-only simulation. 

In Section 2 of this paper, we describe the datasets used for this study, and the methods used to classify galaxies as elliptical or disk. In Section 3 we present the observed group mass-morphology relation, and compare it with that derived from the simulation data. We then discuss the implications of these results and present our conclusions in Section 4.

%=========================================

%2
\section{Methods}

%2.1
\subsection{Data}
\label{Data}

The GAMA survey was carried out at the 3.9m Anglo-Australian Telescope \citep{2011MNRAS.413..971D}. The survey targeted five regions, centred on RA$\sim$ 9h, $\sim$12h and $\sim$15h, with each measuring a size of 12x5 degrees$^{2}$. These regions were chosen to overlap with SDSS coverage, allowing the target galaxies to be selected from optical SDSS images. Here we use the three equatorial regions of the GAMA II dataset, where the detection limit in all three regions is a \citet{1976ApJ...209L...1P} magnitude of $r_{pet}$ < 19.8 mag. 

 One of the significant advancements made by GAMA over previous surveys was its high completeness to close pairs  \citep[e.g.][]{2011MNRAS.416.2640R}. The survey has a spectroscopic completeness level of 98.5 per cent in all three equatorial survey regions \citep{2015MNRAS.452.2087L,2011MNRAS.413..971D}. This very high completeness ensures that, in the majority of cases, galaxy groups can be identified in their entirety up to the magnitude limits of the survey. This allows for a more complete study of small galaxy groups and the accurate determination of their properties such as group virial mass.

\citet{2011MNRAS.416.2640R} created a catalogue of galaxy groups in the GAMA survey. They identified groups using a friends-of-friends algorithm, which takes into account the potential members' proximity in both their projected positions and redshift measurements.

\citet{2011MNRAS.416.2640R} assumed the groups are in a state of virial equilibrium and used the fact that the dynamical mass of a virialised system scales with $\sigma^{2}R$, where $\sigma$ is the velocity dispersion and $R$ the group radius. The group mass is then given by:

\begin{equation}
    \frac{M_{FoF}}{h^{-1}M_{\odot}} = \frac{A}{G / (M_{\odot}^{-1}km^{2}s^{-2}Mpc)}(\frac{\sigma_{FoF}}{km s^{-1}})^{2} \frac{Rad_{FoF}}{h^{-1}Mpc},
\end{equation}

\noindent where $G$ is the gravitational constant, $h$ is the Hubble constant and $Rad_{FoF}$, $\sigma_{FoF}$ and $M_{FoF}$ are the radius, velocity dispersion and mass of the groups respectively. $A$ is a scaling factor whose value is determined by comparing the calculated $M_{FoF}$ with the true mass of the dark matter halos within simulated groups. These dynamical masses of the GAMA groups, which include the gravitational influence of the dominating dark matter, allow for a direct mass comparison with dark matter only simulations. 

We used the G3C Version 7 group catalogue for this paper. We first removed any galaxies without morphological data \citep[those not in the SersicCatAll Version 7 of][see Sec.~\ref{sec-class}]{2012MNRAS.421.1007K}, giving 58,492 galaxies in 19,010 groups. For the determination of the group mass-morphology relation, we selected all groups with redshift $z\leq0.15$ and with 3 or more members, leaving 10,849 galaxies in 2,643 groups. 

%=========================================

%2.2
\subsection{Galaxy Classification}
\label{sec-class}

We used an automated galaxy classification method developed by \citet{2012MNRAS.421.1007K} to classify the GAMA galaxies as one of two morphology classes: ellipticals and disks. \citet{2012MNRAS.421.1007K} used the GALFIT package  to fit a S\'ersic function to each galaxy image, providing the S\'ersic index $n$ which best describes the luminosity profile of the galaxy. The model fitting was carried out for each of the $ugrizYJHK$ bands. Here we use the SersicCatAll Version 7 \citep{2012MNRAS.421.1007K}, along with absolute magnitudes from the catalogue StellarMasses Version 18 \citep{2011MNRAS.418.1587T} datasets. 

\begin{figure}
\includegraphics[width=\columnwidth]{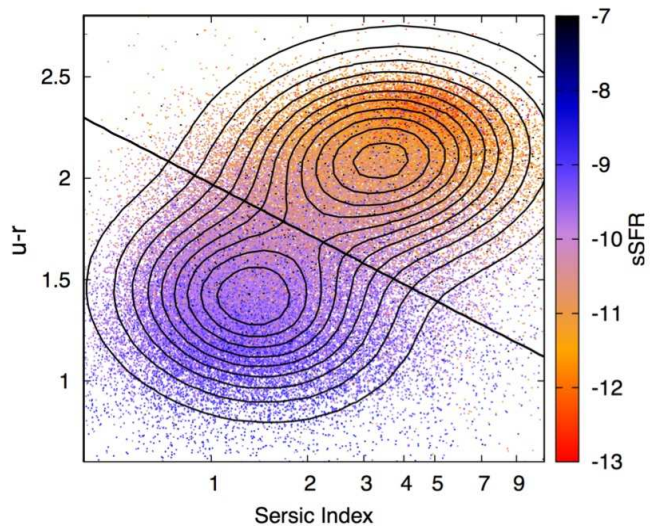}
   \caption{Galaxy classification diagram showing the S\'ersic index and $u-r$ colour index for all galaxies in the GAMA group catalogue. The contours show a two-component Gaussian fit to the distribution. Galaxies are colour-coded by their specific star formation rates. The solid line shows the relation we use to separate he galaxy population into two groups: star-forming disk-type galaxies (lower left) and elliptical galaxies with low star formation rates (upper right) galaxies.}
 \label{Sersic_colour}
\end{figure}

We show the S\'ersic index of galaxies within the GAMA group catalogue plotted against their respective colour indices in Fig.~\ref{Sersic_colour}. This plot separates the catalogue into two distinct populations. The first population is centred on $n = 1.5$ and features a bluer colour index, characteristic of a typical disk galaxy population, while galaxies within the population centred on $n = 4$ typically have redder colour indices as would be expected from a population of elliptical galaxies. The plot has been colour coded with the galaxies' specific star formation rates (sSFR) determined by \citet{2013MNRAS.430.2047H}, revealing that galaxies in the lower-S\'ersic-index population generally have higher sSFRs, as would be expected from a disk galaxy population. 

We performed a two-component Gaussian fit to the distribution (shown by the contours in Fig.~\ref{Sersic_colour}). We then found a line bisecting the two populations at the point of lowest density to divide the distribution into red high-n and blue low-n galaxies \citep{2012MNRAS.421.1007K}. To confirm the reliability of our classification method, we visually classified members of six galaxy groups across the group-mass range and compared these classifications to the automatic method. In particular, we found galaxies whose light was contaminated by (projected) nearby objects are still correctly classified by the automatic method.  

%=========================================

%2.3
\subsection{Simulations} \label{Simulations}

N-body (dark matter-only) simulations have been very successful in tracking the hierarchical formation of structure and the effect of mergers in the mass assembly of galaxies \citep[]{2008MNRAS.384....2G, 2009ApJ...701.2002G} as well as on galaxy morphologies and internal structures \citep[]{1994ApJ...437L..47M, 1996ApJ...465..278J, 2003ApJ...597..893N, 2008ApJ...680..295B}.

We carried out a GADGET 2 dark-matter-only simulation \citep{2005MNRAS.364.1105S} to investigate the merger activity of sub halos within galaxy groups across the group mass range covered by the observational part of this study. The simulation used the following cosmological parameter values: $\Omega_m = 0.307$,  $\Omega_\Lambda = 0.693$, $b = 0.04825$, $h = 0.6777$ and $\sigma_8 = 0.8288$ \citep{2014A&A...571A...1P,2014A&A...571A..16P}. We carried out the simulation 
in a box of size 150 Mpc/h per side with 1024$^{3}$ particles resulting in a particle mass of $2.67898\times 10^{8}M_{\odot}$/h. We identified the halos using the Rockstar Halo Finder \citep{2013ApJ...762..109B} and constructed the halo assembly histories using the Consistent Trees algorithm \citep{2013ApJ...763...18B}. 

We took all sub halos contained within a given host halo to be members of a galaxy group. The mass of the group is then equal to the mass of the host halo i.e. the number of dark matter particles contained within the radius of the host halo. As detailed in Section~\ref{sec-class}, the GAMA group masses determined by \citet{2011MNRAS.416.2640R} are a measure of the total mass of the system including the dark matter halo within which the group is situated. These masses can therefore be compared directly with those of the host halos in the GADGET 2 simulation.

\afterpage{
\begin{figure*}
\Large
\centering
\scalebox{0.65}{\input{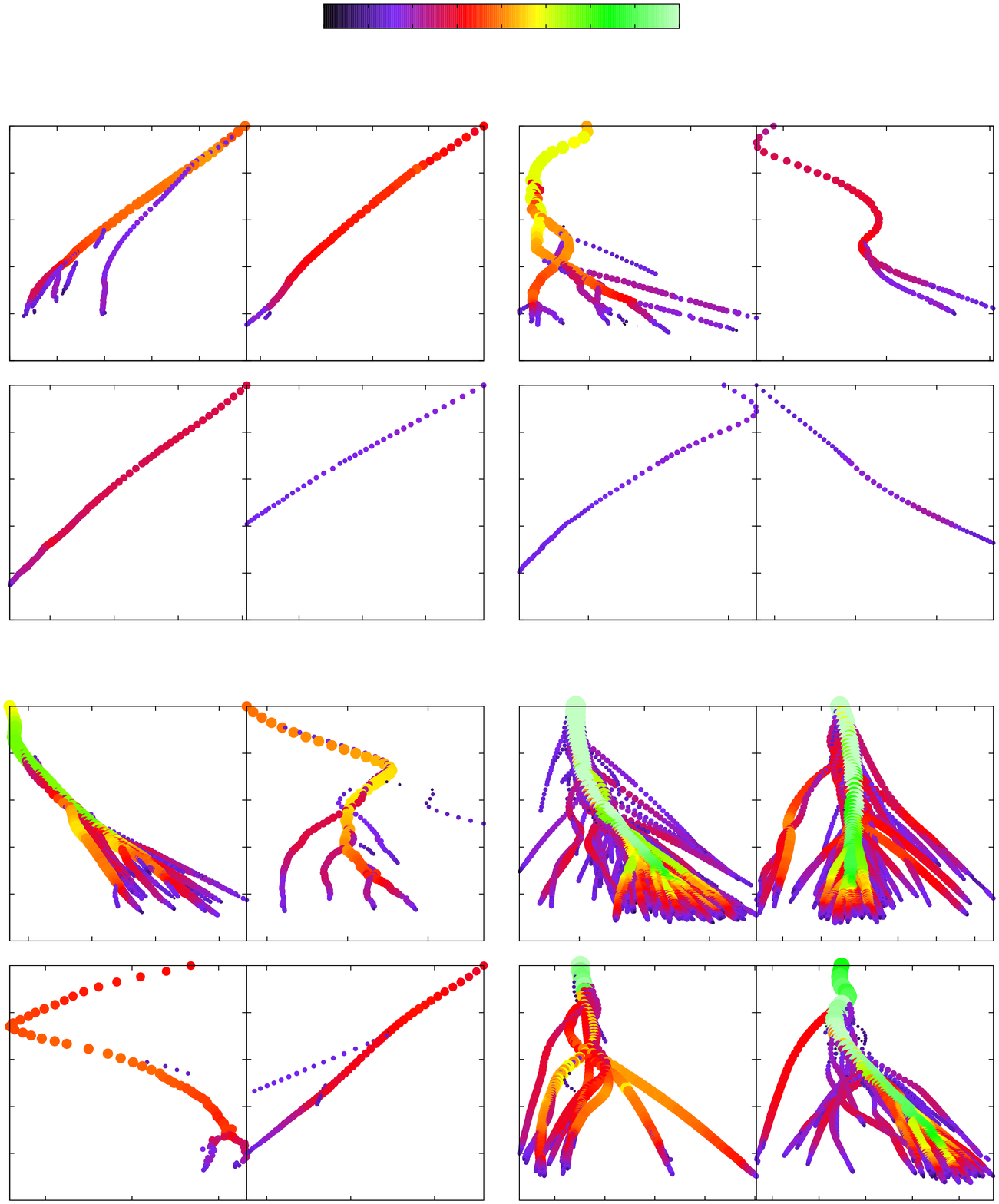}}
 \caption{Comparisons of the merger histories of galaxies in different group environments. For each of 4 groups of increasing mass, the panels show the histories of the 4 largest group members. Each panel plots the co-moving positions (projected on the horizontal axis) of all the precursor halos as a function of the cosmological scale factor, $a(t)$, on the vertical axis. The sequences start at a scale factor of $a=0.1$ and end at the present epoch ($a=1$). The group masses are shown above each set of 4 panels; the group member masses are shown inside each panel. The sizes and colours of the points are scaled by the masses of the precursor halos, indicated by the colour bar. The group members in the more massive environments have much richer merger histories.}
   \label{halo_trees}
\end{figure*}
\clearpage
}

We initially used the major merger events flagged by Rockstar, however the algorithm does not take into account mass loss during the merger process in determining the merger mass ratios, and spurious major merger events were present in the data. We therefore developed our own merger-identifying algorithm which also determined the amount of mass being added to a halo in every merger event.

Assuming a galaxy exists in the centre of each halo (including the host halo), we identified galaxies contained within each group halo and analysed the merger history for each of these galaxies. We designed our algorithm to take into account both major and minor merger events. We need to include minor merger events as \citet{2007A&A...476.1179B} showed that when the cumulative merged mass in a system accreted via minor mergers reaches 30 per cent of the final total mass, the system features structure and dynamics resembling those of an elliptical galaxy.

Halos can begin interacting and transferring mass long before they become one halo in the tree catalogue. In particular, the smaller halo often loses significant mass before the final merger. If the mass ratio is determined at the step before they fully merge, it will then be significantly underestimated. We therefore go through the halo trees and identify when two or more halos merge into a single halo. These halos are then followed back through time, and when they fall outside 1.5 times the virial radius of the main galaxy, their masses at that time are compared to that of the the main galaxy to determine the merger ratio and the merged mass added. If the ratio at that point exceeds 1:3, we flag the event as a major merger. The merged masses of all merger events are added together to find the cumulative merged mass. We identify galaxies in the final epoch of the simulation as elliptical (i.e. transformed) if either they have accumulated over 30 per cent of their final mass from minor mergers, or if they have experienced at least one major merger event.

We estimated how each group in the simulation would appear in the GAMA data using an abundance matching approach. This naturally accounted for the GAMA survey limits, allowing for comparison with the observational results. For each group in the simulation, we randomly selected a group of similar mass in the GAMA catalogue, and noted the number of observed group members, $N_g$. We then identified the largest $N_g$ dark matter halos in the simulated group as the `observed' members of that simulated group. This assumes a one-to-one correspondence between the dark matter halos and observed galaxies for the largest members of each group. This approach gives a distribution of group multiplicity matching the GAMA sample in each group mass bin. We then followed the same procedure as that for the observational analysis, where the galaxies were binned by their host group masses, and the elliptical fraction determined for each mass bin. 

In order to demonstrate the variations in merger activity experienced by galaxies in different environments, we selected several groups across the mass range and plotted the merger histories of the member galaxies. Fig.~\ref{halo_trees} shows the merger histories for the four highest mass members of four groups. In low mass groups (10$^{11}M{\odot}$), there are typically only one or two group members which have experienced merger activity, with the rest remaining undisturbed. Toward medium masses of 10$^{12.5}M{\odot}$, the largest group member experiences a high level of merger activity, with large numbers of minor mergers and 1--4 major mergers. The second largest member also has a significant merger history relative to the other members. In the most massive groups, many members feature a very rich merger history with large numbers of major merger events. 

As this simulation features no baryonic matter, some potentially important physical processes involving baryonic matter in mergers are not included in our model. For example, tidal forces and gravitational torques acting on the gas component can alter the structure and gravitational potential of the interacting and remanent galaxies \citep[e.g.][]{1996ApJ...471..115B}, and energy dissipation from gas can alter the scaling relations and the fundamental plane of remanent spheroid galaxies \citep[e.g.][]{2009ApJ...691.1424H}.

%=========================================

%3
\section{Results}

%3.1
\subsection{Observed Groups}

We present our main observational results in this section: the relationship between the elliptical fraction and the host group mass in the GAMA group catalogue. We then apply the same analysis to mass- and redshift-limited samples to demonstrate the effects of the survey limits. Finally, we look at the spatial distributions of spiral and elliptical galaxies within groups, and how this varies across the group mass range.

%3.1.1
\subsubsection{Elliptical Fraction as a Function of Group Mass}

\begin{figure}
\Large
\centering
\scalebox{.7}{\input{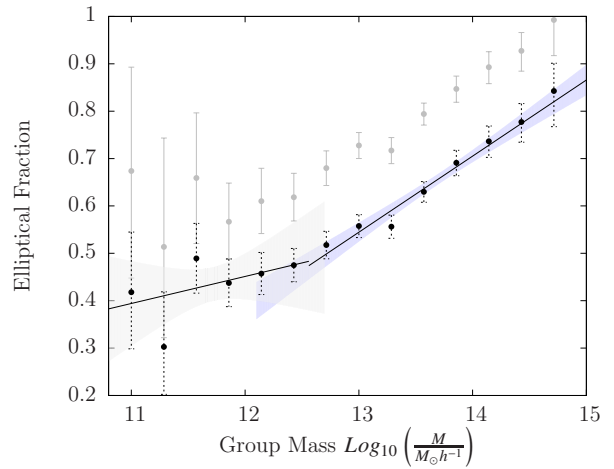}}
 \caption{The observed elliptical galaxy fraction as a function of group mass from the GAMA survey for groups with $NFoF \geq 3$ and $z \leq 0.15$. The error bars denote the 95 per cent binomial confidence intervals in the elliptical fraction. We fitted linear models to the relations for group masses above and below 10$^{12.5}M_{\odot}$; the shaded regions show 95 per cent confidence bounds on the slopes. Grey points show the elliptical fraction for groups with $NFoF \geq 4$, offset vertically by +0.15 for clarity. The elliptical fraction decreases linearly as group mass decreases down to 10$^{12.5}M_{\odot}$, below which the relation flattens out.}
\label{obs_relation}
\end{figure}

Our main observed galaxy group sample consisted of 10,849 galaxies in the 2,643 GAMA groups with $NFoF \geq 3$ and $z \leq 0.15$ (Section~\ref{Data}). We binned the galaxies by the masses of their host groups, and determined the fraction of galaxies automatically classified as elliptical for each bin. We present the elliptical fraction as a function of group mass in Figure~\ref{obs_relation}. The elliptical fraction remains fairly constant from the least-massive groups up to a group mass of ~10$^{12.5}M_{\odot}$. Above this turnover mass, the elliptical fraction increases continuously up to the largest groups in this study, at a rate of 0.161 $\pm$ 0.001 per dex of group mass. We also show in Figure~\ref{obs_relation} the resulting elliptical fractions after increasing the group multiplicity limit to $NFoF \geq 4$ (grey points, offset vertically for clarity), demonstrating that by including groups with only three members, we do not introduce additional bias into the observed relationship.

We fitted both one- and two-component linear fits to the data in Fig.~\ref{obs_relation}. The two-component fit was strongly preferred over the one-component fit according to a Bayesian information criterion test. We also tested the sensitivity of the relationship determined above to small variations in the placement of the bisecting line used for classification (Fig.~\ref{Sersic_colour}). Both the slopes and the turnover point remained fixed, indicating that our findings regarding the relative elliptical fractions across group masses are independent of the precise placement of the dividing line. As an extreme case, separating the sample purely by S\'ersic index (a vertical separation in Fig.~\ref{Sersic_colour}) still gives results similar to that of Fig.~\ref{obs_relation}.

%3.1.2
\subsubsection{Effect of Uncertainty in Group Masses} \label{mass_uncertainties}

We then used Monte Carlo methods to simulate the effect of uncertainties in the group masses on the relation in Fig.~\ref{obs_relation}. Specifically, we considered if a single linear relation could be consistent with the observed flattening below group masses of 10$^{12.5}M_{\odot}$ after including the mass uncertainties. The mass uncertainty could have this systematic effect due to a form of Eddington bias: there are more groups above the transition point than below it, so a random uncertainty in group masses results in more higher mass groups shifting across this point towards lower masses, thereby increasing the elliptical fraction in the lower mass bins. 

\citet{2011MNRAS.416.2640R} estimated the uncertainty in the group masses by applying their group finding techniques to mock simulations with known group masses. The relationship between true and estimated group masses took the shape of an elongated normal distribution twisted away from the 1:1 slope \citep[][Figure 7]{2011MNRAS.416.2640R}. Specifically, groups of a mass below $10^{13}M_{\odot}$ have estimated masses which are on average higher than the true mass, while more massive groups have estimated masses below their true value on average. They determined a group-multiplicity-dependent 1-sigma uncertainty in their derived group masses as:

\begin{equation}
    \log_{10}{\frac{M_{err}}{h^{-1}M_{\odot}} }= 1.0 - 0.43\log_{10}(N_{FoF}),
	\label{eq:quadratic}
\end{equation}

\noindent where $N_{FoF}$ is the number of galaxies in the group and $M_{err}$ is the mass error. Robotham et al. predicted the effects of this uncertainty on the true masses in their simulation using this relation:

\begin{equation}
    \frac{M_{new}}{h^{-1}M_{\odot}} =  \frac{M_{FoF}}{h^{-1}M_{\odot}}10^{G(0,\log_{10}(M_{err}/h^{-1}M_\odot))},
	\label{eq:quadratic2}
\end{equation}

\noindent where $M_{FoF}$ is the group catalogue mass and $G(x,y)$ is a random multiplicative factor taken from a normal distribution with mean $x$ and deviation $y$. This  reproduced the distribution of their estimated group masses, including the twisting away from the 1:1 slope. 

We used these results to predict the effect of mass uncertainty in our own results as follows.

1. We first generated an approximation of the group mass distribution as it would appear without biases introduced by the mass uncertainties. We extracted the mean relationships (and hence the mean conversion factors) between the calculated and true group masses for each multiplicity range presented by \citet[][Figure 7]{2011MNRAS.416.2640R} using the major axes of the overlaid 10 and 50 percentile contours. For each GAMA group, depending on its multiplicity, we then altered its mass by the appropriate conversion factor, producing the desired group mass distribution. 

\begin{figure}
    \large
\centering
\scalebox{0.8}{\input{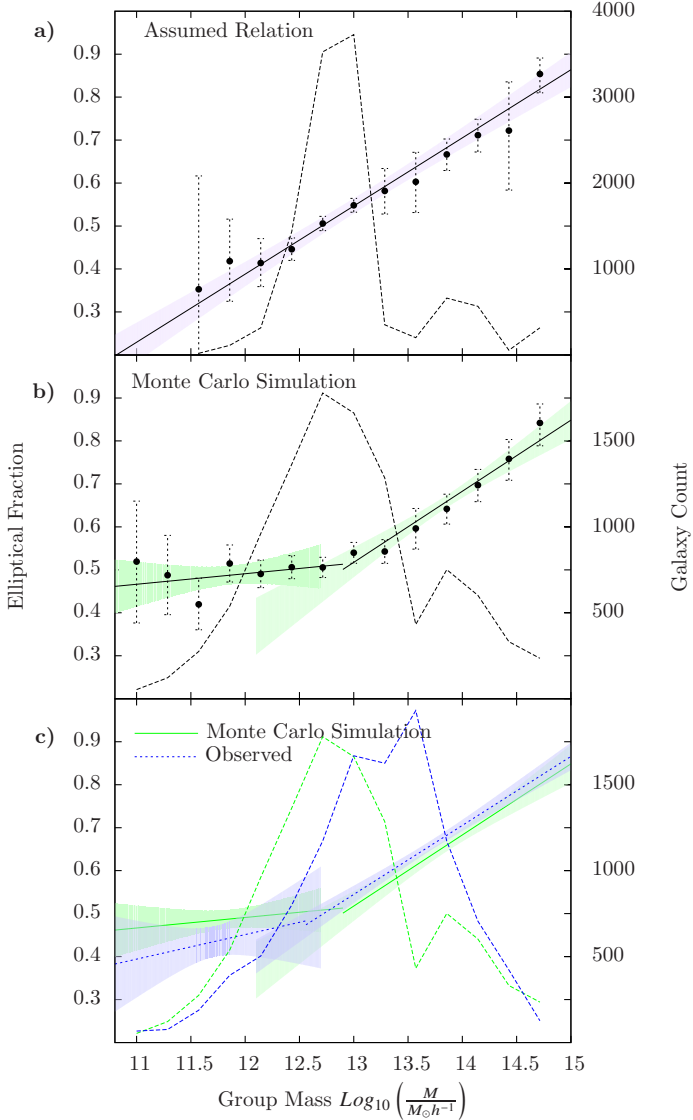}}
    \caption{The effect of uncertainty in group masses on the group mass morphology relation. a) The elliptical galaxy fraction for one simulation of our survey assuming a constant linear relation over the whole range of group mass. No uncertainties have been added to the group masses. b) The resulting group mass-morphology relation after adding uncertainties to the group masses. c) The group mass-morphology relation from b) (green, solid lines) overlaid by the observed relation found in Fig.~\ref{obs_relation} (blue, dotted lines). Dotted curves show the number of galaxies in each bin. Shaded areas are the 95 per cent confidence bounds for the linear regressions. The re-introduction of mass uncertainties to a) reproduces the flattening seen below $10^{12.5}M_{\odot}$ in the observational result while maintaining the original gradient at higher masses.}
    \label{monte_carlo}
\end{figure}

2. We then assumed that the relationship found above 10$^{12.5}M_{\odot}$ in Fig.~\ref{obs_relation} extends across the full group mass range. Based on this assumption, the expected mass-dependent elliptical fraction was calculated for each group. Galaxies within each group were then assigned a random number between zero and one. They were classified as a disk if this number was above the expected fraction for their host group's mass, or an elliptical otherwise. An example of the resulting group mass-morphology relation is shown in Fig.~\ref{monte_carlo}a. 
  
3. We then re-introduced random uncertainties to the group masses using Equation 3 and found the group mass-morphology relation as before. Fig.~\ref{monte_carlo}b shows the group mass-morphology relation after applying these mass uncertainties to Fig.~\ref{monte_carlo}a. The resulting relation appears very similar to our observed result in Fig.~\ref{obs_relation}, with a clear flattening of the relation below group masses of 10$^{12.5}M_{\odot}$.   Fig.~\ref{monte_carlo}c shows the regression of Fig.~\ref{monte_carlo}b (green shaded area, dotted line) overlaid by the observational result from Fig.~\ref{obs_relation} (blue shaded area, solid line). The gradient of the relation resulting from the Monte Carlo simulation in the high mass regime is $0.165\pm0.017$ per dex of group mass, remains in good agreement with the observational result of 0.163$\pm$0.012 per dex of group mass. For the low mass regime < 10$^{12.5}M_{\odot}$, the Monte Carlo simulation gives a gradient of $0.025\pm0.020$ per dex of group mass, consistent with that found from the observational result, $0.057\pm0.039$ per dex of group mass. 

4. We repeated the Monte Carlo simulation 5000 times to measure the spread in the results. Below and above 10$^{12.5}M_{\odot}$ the resulting calculated gradients are  0.025 (-0.01, +0.070) and 0.163 (-0.118, +0.210) per dex of group mass respectively, where the uncertainty ranges give the interval containing 95 per cent of the calculated gradients.

This illustrates that while the group mass uncertainties have little impact on the relationship between elliptical fraction and group mass in the higher mass regime, the uncertainties result in a significant flattening in the relationship at lower masses. Our observed results are therefore consistent with a linear relationship between elliptical fraction and group mass across the entire mass range. 

%3.1.3
\subsubsection{Effect of Survey Limits} \label{Survey_Limits}

\begin{figure*}   
	\centering
	\includegraphics[width=0.8\linewidth]{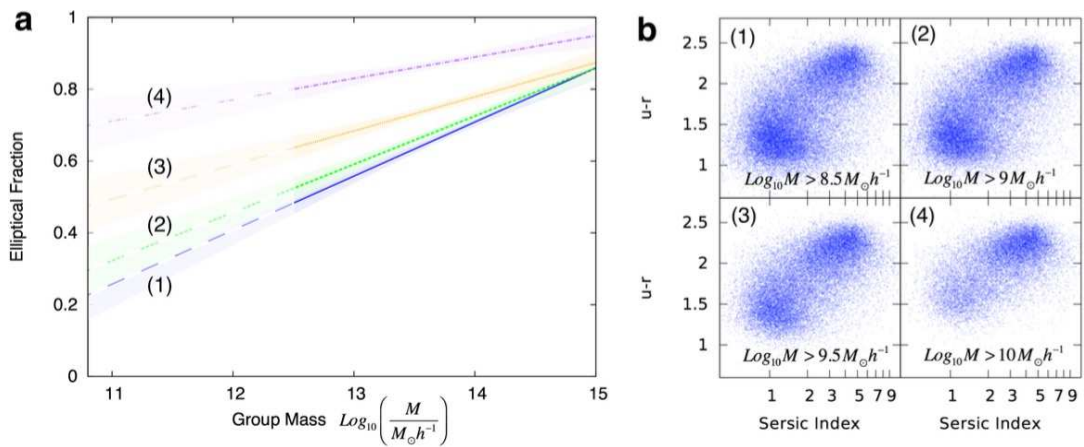}
    \caption{The effect of galaxy mass limit on the observed group mass morphology relation. Panel a) shows the observed relations for four samples with increasing lower limits to the galaxy masses as indicated in b). The lines of best fit are calculated for group masses above 10$^{12.5}M_{\odot}$, all for the main sample with $z \leq 0.15$. Long dashed segments indicate extensions of the slopes to lower group masses. Increasing the lower mass limit raises the elliptical fraction and reduces the gradient. Panel b) shows the S\'ersic index-colour plots as in Fig.~\ref{Sersic_colour}, demonstrating that as the galaxy mass limit is raised, galaxies are removed from the lower left (disk) galaxy population while the upper right elliptical population remains relatively unaffected.}
      \label{vary_m}
\end{figure*}

\begin{figure*}
	\centering
	\includegraphics[width=0.8\linewidth]{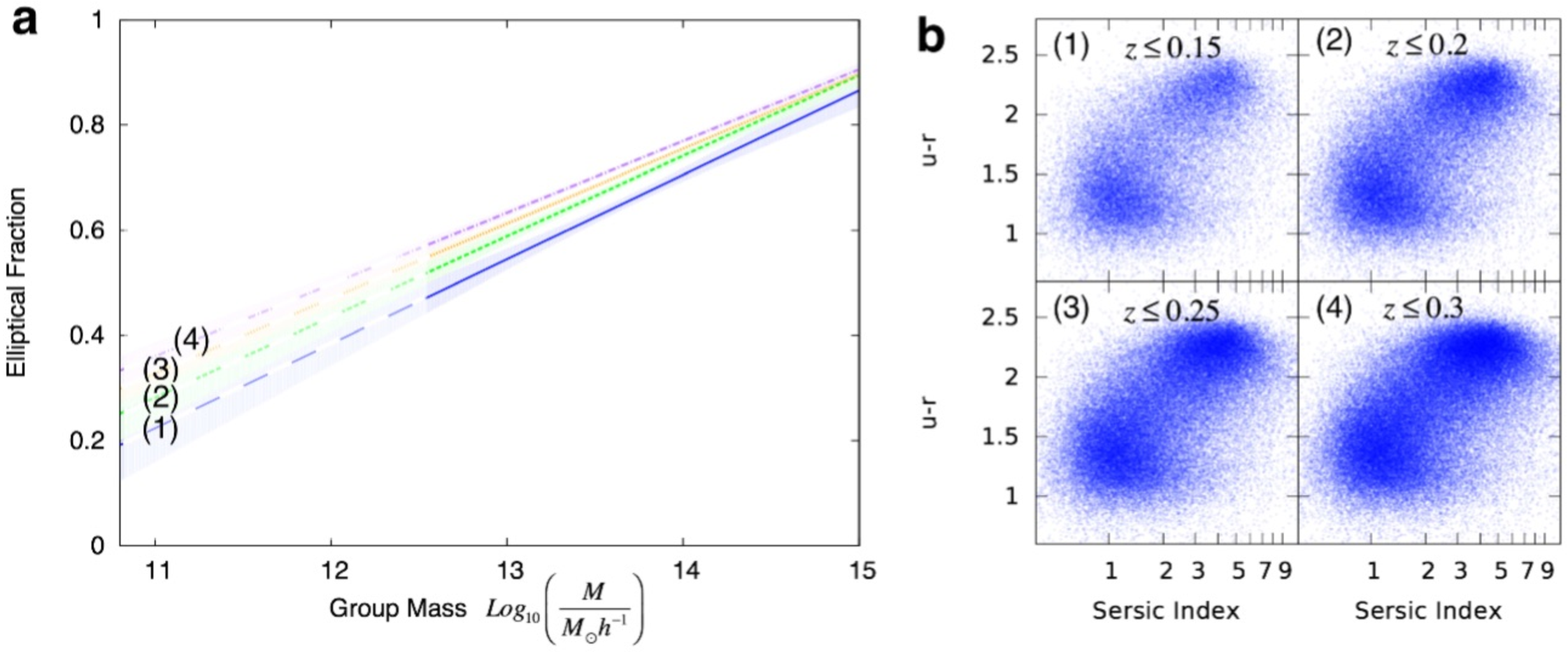}
   \caption{The effect of redshift limit on the observed group mass morphology relation. Panel a) shows the observed relations for four samples with increasing upper redshift limit, as indicated in b). The lines of best fit are calculated for group masses above 10$^{12.5}M_{\odot}$.  Long dashed segments indicate extensions of the slopes to lower group masses. Increasing the upper redshift limit increases the elliptical fraction and slightly reduces the gradient. Panel b) illustrates how the galaxy population changes in each redshift-limited sample, showing that as the redshift limit is increased, more ellipticals are added to the population than disks.}
        \label{vary_z}
\end{figure*}

The magnitude limits of the GAMA survey introduce a potential source of bias into these results. With increasing group redshift, the smaller, fainter disk galaxies begin to fall below the detection limits while the brighter, high mass elliptical galaxies remain within the survey limits. At even higher redshifts, all members of the smaller groups fall below the detection threshold, leaving only the largest galaxies in the largest groups within the survey limits. In the analysis of the GAMA data in Fig.~\ref{obs_relation}, we restricted the sample to $z \leq 0.15$ and placed no lower limit on the galaxy mass. To separate the impact on the determined group mass-morphology relation of both the increasing galaxy lower-mass limits with increasing redshift and the effects of increasing the redshift limit alone, we analysed samples with varying redshift and galaxy mass limits.  

We first isolated the impact of varying the lower galaxy mass limit in the sample, with the redshift limit held fixed at $z \leq 0.15$. The resulting gradients for lower limits of M $\geq 10^{8.5}$, $10^{9}$, $10^{9.5}$ and $10^{10} M_{\odot}$ are shown in Fig.~\ref{vary_m}a. As the lower galaxy mass limit is raised, the elliptical fraction increases across all group masses and the gradients clearly become flatter, with the differences between them becoming more significant as the limit is raised. The distributions of S\'ersic index against colour index in Fig.~\ref{vary_m}b confirm that as the mass limit is raised, the galaxies falling out of the sample are those in the bottom left population, corresponding to blue spirals. 

We then considered the effect of increasing the upper redshift limit (with no imposed galaxy mass limit). Fig.~~\ref{vary_z}a shows the resulting gradients for samples of $z$ $\leq$ 0.15, 0.20, 0.25 and 0.30, along with the S\'ersic index-colour index plots for each sample in Fig.~\ref{vary_z}b. As the upper redshift limit is increased, the gradient again becomes flatter (however not dramatically so) and the overall elliptical fraction increases. Fig.~\ref{vary_z}b illustrates how the upper right populations in the S\'ersic index-colour index plots, corresponding to red ellipticals, increase more quickly than the lower left spiral population as the $z$ limit increases. However, for all redshift-limited samples considered, a strong positive relationship of elliptical fraction with group mass persists. 

Figs.~\ref{vary_m} and ~\ref{vary_z} illustrate that the slope in the group mass-morphology relation is affected by the sample selection limits. Spirals have a lower mass limit relative to ellipticals, as expected, and are therefore the first to fall out of the survey limits as redshift increases. We therefore used abundance matching to tune our simulation to match the properties of the magnitude-limited GAMA group sample as closely as possible.

%3.1.4
\subsubsection{Galaxy Distributions in Groups} \label{obs_distributions}

\begin{figure} 
\Large
\centering
\scalebox{.6}{\input{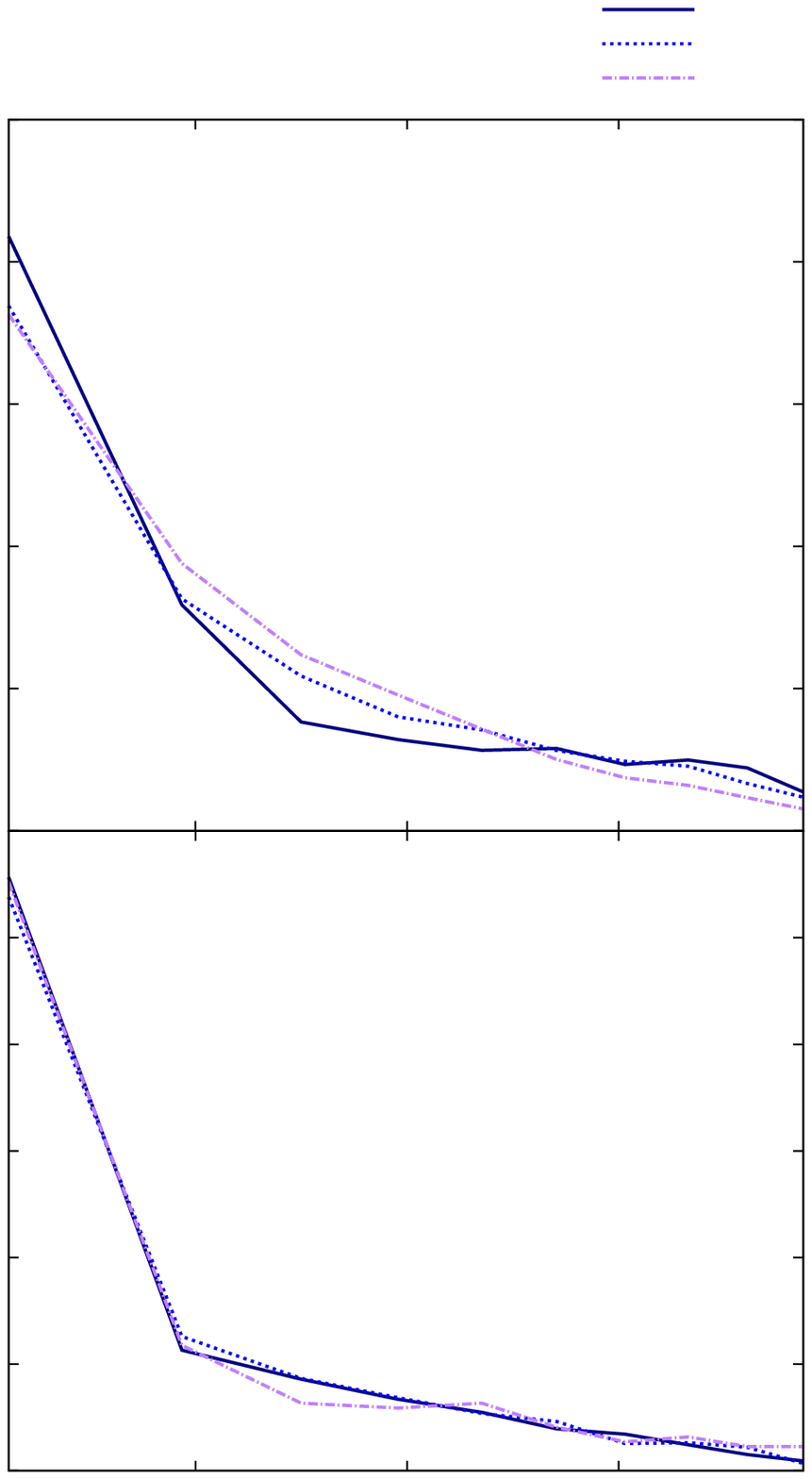}}
    \caption{Radial distribution of galaxies for three group mass bins ($10^{12},10^{13},10^{14}M_{\odot}$, solid, dashed and dot-dashed lines respectively) for the observed GAMA groups (top) and simulation groups (bottom). The radii in each group are all normalised to the radius containing all the group members. The galaxy distributions are relatively similar between the observed and simulated groups.}
    \label{count_distributions}
\end{figure}

\begin{figure} 
\Large
\centering
\scalebox{.6}{\input{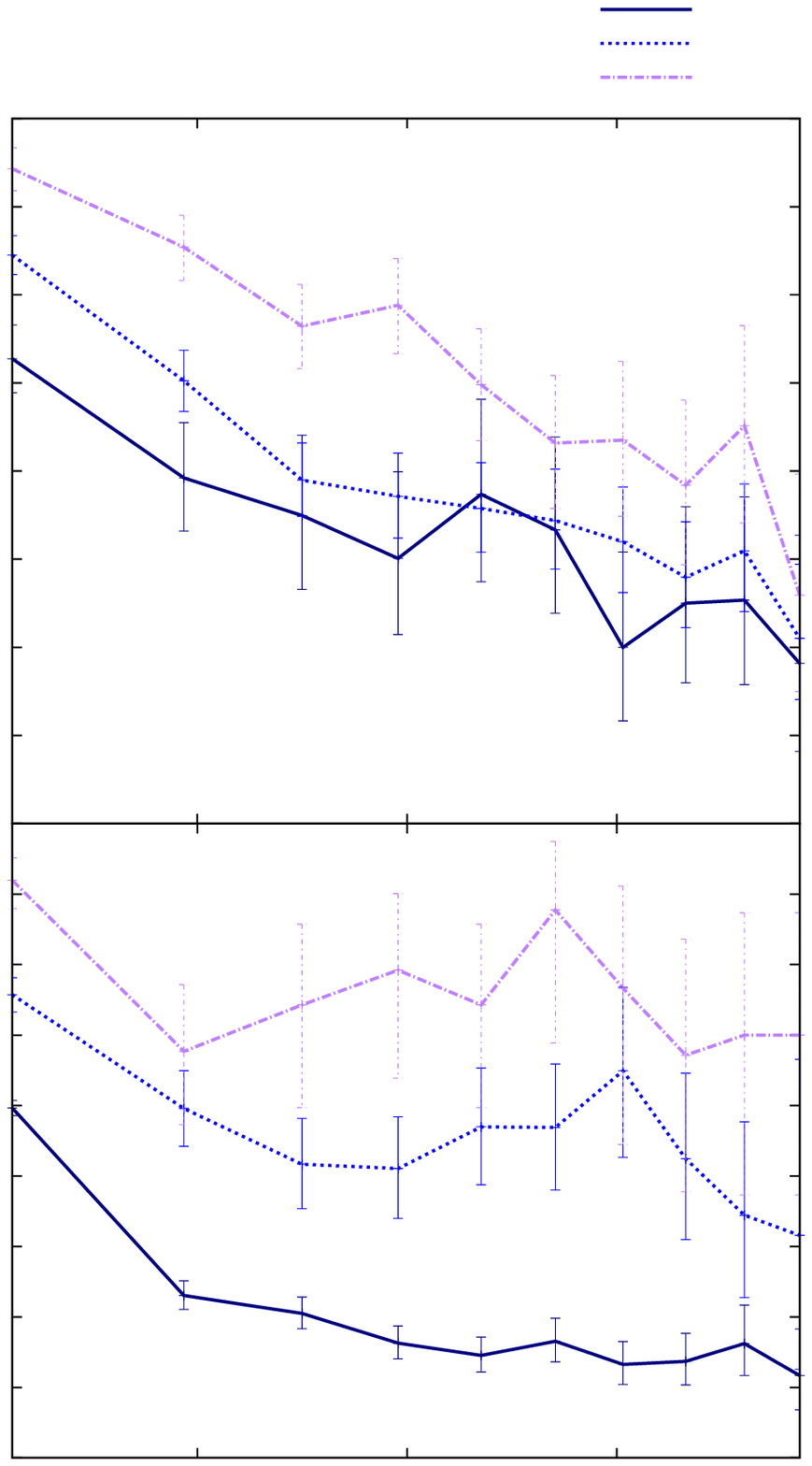}}
    \caption{The elliptical fraction for three group mass bins ($10^{12},10^{13},10^{14}M_{\odot}$, solid, dashed and dot-dashed lines respectively) for the observed GAMA groups (top) and simulation groups (bottom). The radii in each group are all normalised to the radius containing all the group members. The error bars denote the 95 per cent binomial confidence intervals. The elliptical fraction peaks at the centre in both observed and simulated groups, .}
    \label{ratio_distributions}
\end{figure}

The relative spatial distributions of the two galaxy classes in the groups and how this varies with increasing group mass can illuminate where in the groups the transformations are occurring. Within three group-mass bins centred on $10^{12}$, $10^{13}$ and $10^{14}M_{\odot}$  each, the galaxies were further binned according to their distance from the group centre normalised by their group's radius (here defined as the radius containing all galaxy members), with the radius range of each bin chosen such that the bins have equal area. The resulting distributions of the total galaxy populations and the elliptical fraction are plotted in the upper panels of Fig.~\ref{count_distributions} and Fig.~\ref{ratio_distributions} respectively. In all three group-mass bins, the galaxy distributions are very similar and strongly peaked towards the group centre, and the elliptical fraction is highest in the centre, decreasing steadily with increasing radius. The rate of decrease across the three group-mass bins are relatively similar, indicating that the increasing transformation rate of galaxies with increasing group masses is occurring uniformly throughout the group.

%3.2
\subsection{Simulated Groups}

We determined the extent to which the observed transformation of spirals to ellipticals (Fig.~~\ref{obs_relation}) can be explained by merger events by identifying galaxies with extensive merger histories as `elliptical' in our simulated groups, as described in Sec.~\ref{Simulations}. First we derived an equivalent group-mass morphology relation from merger activity in the simulation and compared these results with the GAMA data. Second, we measured the intra-group distributions of galaxies in the simulation and also compared these to the GAMA data.  

%3.2.1
\subsubsection{Merger Events}
\label{Merger_Events}

\begin{figure} 
\Large
\centering
\scalebox{.6}{\input{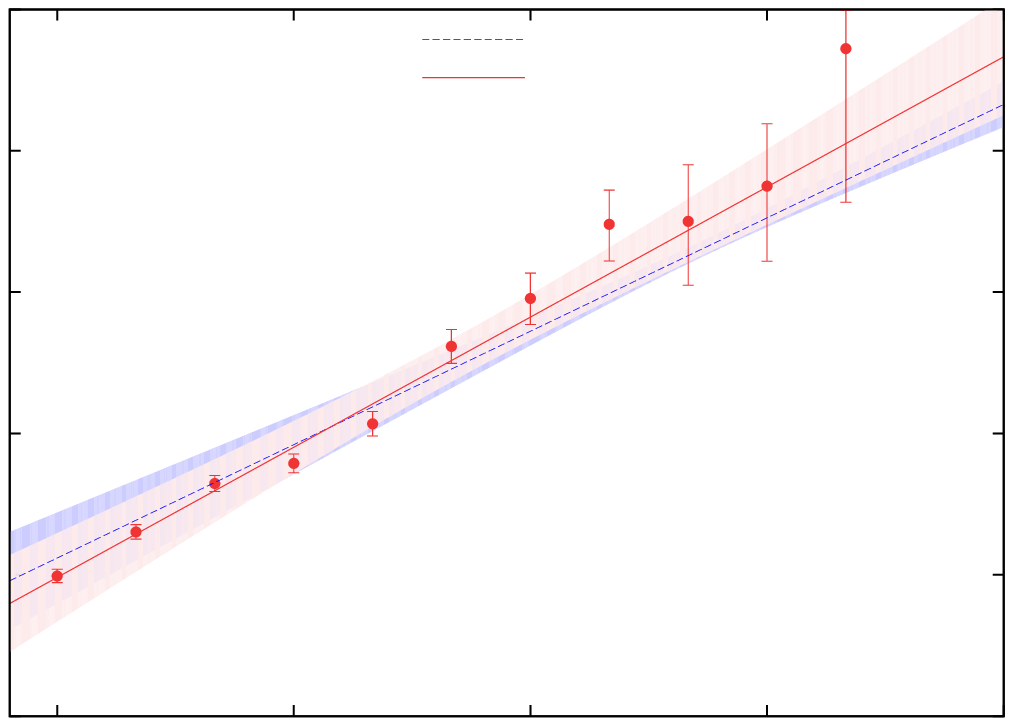}}
 \caption{The group mass-morphology relation (red points and solid red line) predicted by our cosmological simulation. We classify each galaxy in the simulation as elliptical if it either underwent a major merger event or if its cumulative merged mass exceeded 30 per cent (see Sec.~\ref{Simulations}). The blue dotted line shows the likely relation underlying the observational results (see Sec.~\ref{mass_uncertainties}). The error bars show the 95 per cent binomial confidence intervals in the elliptical fractions, and the shaded regions show the 95 per cent confidence interval of the linear regression fits. The group mass-morphology relation derived from the simulations is in close agreement with the observational result.} 
 \label{simulation_relation_2}
\end{figure}

We present the group mass-morphology relation for the simulated groups as the red solid line in Fig.~\ref{simulation_relation_2}, with the regression from the observed relation above 10$^{12.5}M_{\odot}$ overlaid (blue dotted line). The group mass-morphology relation from the simulation, a fraction of $0.184 \pm 0.001$ per cent per dex of group mass, is consistent with the observational value (for higher masses) of $0.161 \pm 0.001$ per dex of group mass. As there are no uncertainties in the simulated group masses and hence no Eddington bias at play, the relation seen in the simulation data holds through all group masses down to  10$^{11}M_{\odot}$. 

To test if the resulting merger rates identified in the simulation are reasonable, we compared our calculated merger rates with those inferred from observations. \citet{2014ApJ...795..157K} determined from observations of close galaxy pairs that a typical galaxy of stellar mass $10^{10.7} M_{\odot}$ experienced between 0.2 and 0.8 major merger events since $z = 1$, depending on merger timescales and the fraction of close pairs which eventually merge. 
In our simulations, the corresponding dark matter halo mass for galaxies is about $10^{13} M_{\odot}$ (e.g. using the relation of \citet{2015MNRAS.454.1161Z}). Adjusting our definition of a major merger to one where the merger ratio exceeds $1:10^{0.4}$ as per \citet{2014ApJ...795..157K}, our simulations predict that a galaxy of halo mass $10^{13} M_{\odot}$  has undergone an average of $0.12\pm0.11$ major merger events since $z = 1$. This is just below the lower limit of the range (0.2 to 0.8) determined by \citet{2014ApJ...795..157K}. We conclude that there is not a strong inconsistency with the observed merger rates, but we reserve a detailed comparison for future work.

%3.2.2
\subsubsection{Galaxy Distributions}

Figs.~\ref{count_distributions} and \ref{ratio_distributions}  show the galaxy distributions and the elliptical fraction as a function of radius for the simulation groups (bottom row) and the observational results (top row, see Section~\ref{obs_distributions}). The groups were placed into three mass bins centred on $10^{12},10^{13},10^{14}M_{\odot}$, then the galaxies were placed into normalised equal-area radial bins as before. The distributions of galaxies in the simulation groups are similar to the observed groups (Fig.~\ref{count_distributions}), with a slightly stronger central peak. The elliptical fractions are again peaked in the group centres as also seen the the GAMA groups (Fig.~\ref{ratio_distributions}), with similar averaged declines towards the edge albeit with more variations, particularly for the largest groups. The generally higher elliptical fractions seen in the $10^{12}M_{\odot}$ mass bin of the observed groups compared to the simulation groups are a byproduct of the mass uncertainty effects discussed in Section~\ref{Survey_Limits}.

%=========================================

%4
\section{Discussion and Summary}

Our observational results reveal a continuous decrease in the elliptical galaxy fraction in groups as the group mass decreases from 10$^{15}M_{\odot}$ to 10$^{12.5}M_{\odot}$. The flattening of the relationship seen below group masses of 10$^{12.5}M_{\odot}$ is most likely caused by a bias introduced by group mass uncertainties, where there are more high-mass groups being erroneously placed in lower mass bins than vice-versa, hence raising the average elliptical fraction in the low-mass bins. Modelling these uncertainties, we found that the observational results are consistent with a continuous decrease in elliptical fraction down to group masses of $10^{11}M_\odot$. We tested the effect of the limits of the GAMA survey on these results and found that the form of the group mass-morphology relation is sensitive to the galaxy mass limit and the sample redshift limit. Changing these limits alters the slope of the relation, but the overall trend is unchanged.

This observational result differs from previous studies which failed to detect any change in elliptical fraction as a function of group mass \citep{2009MNRAS.393.1324B,2012MNRAS.423.3478H}. These previous studies were based on the \citet{2005AJ....130..968M} group catalogue derived from the Sloan Digital Sky Survey: it was only complete for higher galaxy and group masses (compared to the GAMA data used in this paper): our analysis of the effect of such limits in Sec.~\ref{Survey_Limits} demonstrates that raising either of these limits will flatten the observed group-mass morphology relation. A further advantage of the GAMA group catalogue is that the spectroscopic GAMA data are much more complete in crowded group fields than SDSS, resulting in more accurate group detection and measurement.

The observed group mass-morphology relation strongly indicates that one or more processes dependent on the group mass are driving the transformation of galaxy morphologies from disk to elliptical in these groups. We found the elliptical fraction to be higher in the centres of groups relative to the outer edges, indicating that the transformation process is occurring more frequently within the higher density regions of the groups. This observation is consistent with the hypothesis that morphology transformation is being driven by merger activity, which would be expected to increase with density. In apparent contrast to our work,  \citet{Kafle2016} measured the distribution of galaxy masses in the more massive (above $10^{12}M_\odot$) GAMA groups and found no evidence for any change of average galaxy mass with radius from the group centres. We would expect the masses to increase towards the group centres from our results, but we have combined data from a much larger range of environments and used a different sample definition. \citet{Alpaslan2016} measured galaxies in filaments and find that the masses do increase towards the cores of the filaments.

We tested the effect of merger activity by simulating populations of similar mass groups in a dark matter simulation. We predicted the elliptical fraction in the simulated groups by assuming that galaxies which experience major mergers (or have accumulated more than 30 per cent of their final mass in minor mergers) will be transformed to ellipticals. The resulting relation had a slope consistent with relation we measured for the observed galaxies in the GAMA group catalogue. This suggests that merger activity is a major driver of galaxy evolution within galaxy groups, as the increasing merger rate in higher mass simulated groups matches the higher elliptical fraction in high mass observed groups. The spatial distribution of elliptical galaxies in the simulations was also in agreement with the observed distributions, so the observed spatial distributions can also be explained by merger activity, which increases in the higher density group centres. Our simulations did not include baryonic aspects of the merger process as we note in Sec.~\ref{Simulations}, but the general formation of elliptical galaxies by the merging of smaller disk-like galaxies is a very well-established model \citep[as reviewed by][]{2013ApJ...778...61T}.

What is new about the work in this paper is that it is the first to combine a complete group sample spanning such a large range of group masses with cosmological simulations of the same range of group masses. There may be some tension between the merger rates in our simulations and observed merger rate estimates as we discuss in Sec.~\ref{Merger_Events}. If we have underestimated merger rates in our simulations, then a correction would lead to larger predicted elliptical fractions from the simulations, although the correction would need to be modelled as a function of galaxy or group mass. It would also help to improve observational measures by using large integral-field galaxy surveys \citep{2002MNRAS.329..513D,2011MNRAS.413..813C,2015MNRAS.447.2857B,2015IAUS..309...21B} to measure large samples of galaxies in groups for direct kinematic tracers of merger activity. These integral-field surveys could identify fast- and slow-rotating elliptical galaxies in the groups, which should be produced by different merger sequences according to \citet{2014MNRAS.444.1475M}.

In summary, we observed a continuous decrease in the elliptical galaxy fraction in groups with decreasing group mass from 10$^{15}M_{\odot}$ to 10$^{12.5}M_{\odot}$. When we allow for uncertainties in our measured group masses, the data are consistent with a single linear relation over the whole range of group masses observed, from 10$^{11}M_{\odot}$  to 10$^{15}M_{\odot}$, with the elliptical fraction increasing at a rate of $0.16 \pm 0.01$ per dex of group mass. This indicates that the group environment has a significant impact on the rate of galaxy disk-to-elliptical morphology transformations. We measured the rate of merger activity in simulated groups of the same masses and found that the fraction of galaxies which experienced major merger activity in the simulations increases in the same way with group mass as the elliptical fraction in the observed groups, suggesting that the main process responsible for this group-mass-dependent transformation rate is the merger activity experienced by galaxies in these environments.

\section*{Acknowledgements}
This research was conducted by the Australian Research Council Centre of Excellence for All-sky Astrophysics (CAASTRO), through project number CE110001020. SB acknowledges funding support from the Australian Research Council through a Future Fellowship (FT140101166). GAMA is a joint European-Australasian project based around a spectroscopic campaign using the AAT. The GAMA input catalogue is based on data taken from the SDSS and the UKIRT Infrared Deep Sky Survey. Complementary imaging of the GAMA regions is being obtained by a number of independent survey programs including GALEX MIS, VST KIDS, VISTA VIKING, WISE, Herschel-ATLAS, GMRT and ASKAP providing UV to radio coverage. GAMA is funded by the STFC (UK), the ARC (Australia), the AAO and the participating institutions. The GAMA web site is http://www.gama-survey.org/.

%%%%%%%%%%%%%%%%%%%%%%%%%%%%%%%%%%%%%%%%%%%%%%%%%%


\bibliographystyle{mnras}

%%%%%%%%%%%%%%%%%%%%%%%%%%%%%%%%%%%%%%%%%%%%%%%%%%

%%%%%%%%%%%%%%%%% APPENDICES %%%%%%%%%%%%%%%%%%%%%

%\appendix


%%%%%%%%%%%%%%%%%%%%%%%%%%%%%%%%%%%%%%%%%%%%%%%%%%

% Don't change these lines
\bsp	% typesetting comment
\label{lastpage}
\end{document}